\documentstyle[prl,aps,multicol,psfig]{revtex}
\begin{document}
\draft
\title{Quantization of the Hall conductivity well beyond the
adiabatic limit in pulsed magnetic fields}
\author{V.~T.~Dolgopolov and A.~A.~Shashkin}
\address{Institute of Solid State Physics, Chernogolovka, Moscow
District 142432, Russia}
\author{J.~M.~Broto, H.~Rakoto, and S.~Askenazy}
\address{SNCMP INSA 135 avenue de Rangueil 31077 Toulouse cedex, 4,
France}
\maketitle
\begin{abstract}
We measure the Hall conductivity, $\sigma_{xy}$, on a Corbino
geometry sample of a high-mobility AlGaAs/GaAs heterostructure in a
pulsed magnetic field. At a bath temperature about 80~mK, we observe
well expressed plateaux in $\sigma_{xy}$ at integer filling factors.
In the pulsed magnetic field, the Laughlin condition of the phase
coherence of the electron wave functions is strongly violated and,
hence, is not crucial for $\sigma_{xy}$ quantization.
\end{abstract}
\pacs{PACS numbers: 72.20 My, 73.40 Kp}
\begin{multicols}{2}

On recognizing the crucial role of the edge channels in
two-dimensional (2D) electron transport in a quantizing magnetic
field \cite{halp,buttik}, it became pretty clear that the
quantization of the Hall resistance, $R_{xy}$, in Hall bar samples,
which corresponds to the quantum Hall effect \cite{prange}, is not
directly connected with that of the Hall conductivity, $\sigma_{xy}$.
Even if the longitudinal resistance, $R_{xx}$, is negligible, the
measured resistance tensor cannot be converted into the conductivity
one: the net Hall current is a sum of the bulk and edge currents
while the conductivities $\sigma_{xx}$ and $\sigma_{xy}$ are related
to the bulk of the 2D system. Therefore, the conductivity tensor and
the accuracy of $\sigma_{xy}$ quantization should be investigated
independently using the Corbino geometry which allows separation of
the bulk contribution to the measured current. Such an arrangement
was described in the Laughlin \cite{laugh} and Widom-Clark
\cite{widom} gedanken experiments. A (Hall) charge transfer below the
Fermi level between the coasts of a Corbino sample is induced by
magnetic field sweep and thus the shunting effect of the edge
currents is completely excluded. The concept of Ref.~\cite{laugh}
based on gauge invariance leads to the conclusion that at integer
filling factor the conductivity $\sigma_{xy}$ will be quantized if
the magnetic field, $B$, is changed adiabatically so as to keep the
phase coherence of the wave functions on the sample size. The
quantization of $\sigma_{xy}$ follows from the fact that an integer
number of electrons is transferred between the ring edges if the
magnetic flux changes by one quantum. It is clear that the phase
coherence should be the case at field sweep rates when the magnetic
flux change, $\Delta\Phi=\tau L^2{\rm d}B/{\rm d}t$, in a sample with
size $L$ within the settling time, $\tau$, of the wave function phase
is small compared to the flux quantum, $h/e$:

\begin{equation}
{\rm d}B/{\rm d}t < 2\pi\Omega_cB(l/L)^4,\label{eq1}\end{equation}
where $\Omega_c$ is the cyclotron frequency, $l$ is the magnetic
length, and the phase settling time is estimated as the ratio of the
sample size and the phase velocity of an electron,
$\tau=L^2/l^2\Omega_c$.

Doubts about the correctness of the gauge invariance approach were
expressed in Ref.~\cite{riess} and were thought to be supported by
results of the microwave studies, e.g., of Ref.~\cite{kuchar}. In
fact, those studies as well as edge magnetoplasmon \cite{gov} and
related \cite{yahel} experiments are not free of edge current
contribution so that they do not yield the pure $\sigma_{xy}$ and
cannot be an argument against the approach of Ref.~\cite{laugh}. The
value $\sigma_{xy}$ can be measured in the arrangement of the above
gedanken experiments which was employed in the work of
Refs.~\cite{dol,watts,wiegers}. Plateaux with the quantized values of
$\sigma_{xy}$ were indeed observed in the quasistatic measurements of
Ref.~\cite{dol}, even though the quantization accuracy was about 1\%.

Here, we study the charge transfer in a Corbino geometry sample
subjected to a pulsed magnetic field with sweep rate up to $5\times
10^2$~T/s at low temperatures. The conductivity $\sigma_{xy}$ has
been found to be quantized at integer filling factor. This result is
very similar to the data obtained in quasistatically changing
magnetic fields, although at such high sweep rates of the pulsed
magnetic field, the phase coherence of the electron wave functions is
strongly broken. So, the condition of adiabaticity is sufficient but
not necessary for $\sigma_{xy}$ to be quantized.

The samples are Corbino disks fabricated from two wafers of
AlGaAs/GaAs heterostructures containing a 2D electron gas with
mobility $1.2\times 10^6$ and $4\times 10^5$~cm$^2$/Vs at 4.2~K and
density $3.6\times 10^{11}$ and $3.2\times 10^{11}$~cm$^{-2}$,
respectively. Each sample has a circular gate covering a part of the
sample area so that the gated region of the 2D electron system is
separated from the contacts by guarding rings, see Fig.~\ref{f1}. The
radii of the Corbino ring are $r_1$ and $r_2$, and the circular gate
is restricted by radii $r_{1g}$ and $r_{2g}$; two sets of the radii
are employed: (i) $r_1=0.2$~mm, $r_2=0.5$~mm, $r_{1g}=0.305$~mm, and
$r_{2g}=0.39$~mm and (ii) $r_1=1.013$~mm, $r_2=1.119$~mm,
$r_{1g}=1.025$~mm, and $r_{2g}=1.108$~mm. The sample is placed into
the mixing chamber of a dilution refrigerator with a base temperature
of 80~mK. In each pulse, the magnetic field sweeps up to 34~T (or
\vbox{
\vspace{16mm}
\hbox{
\hspace{-0.2in}
\psfig{file=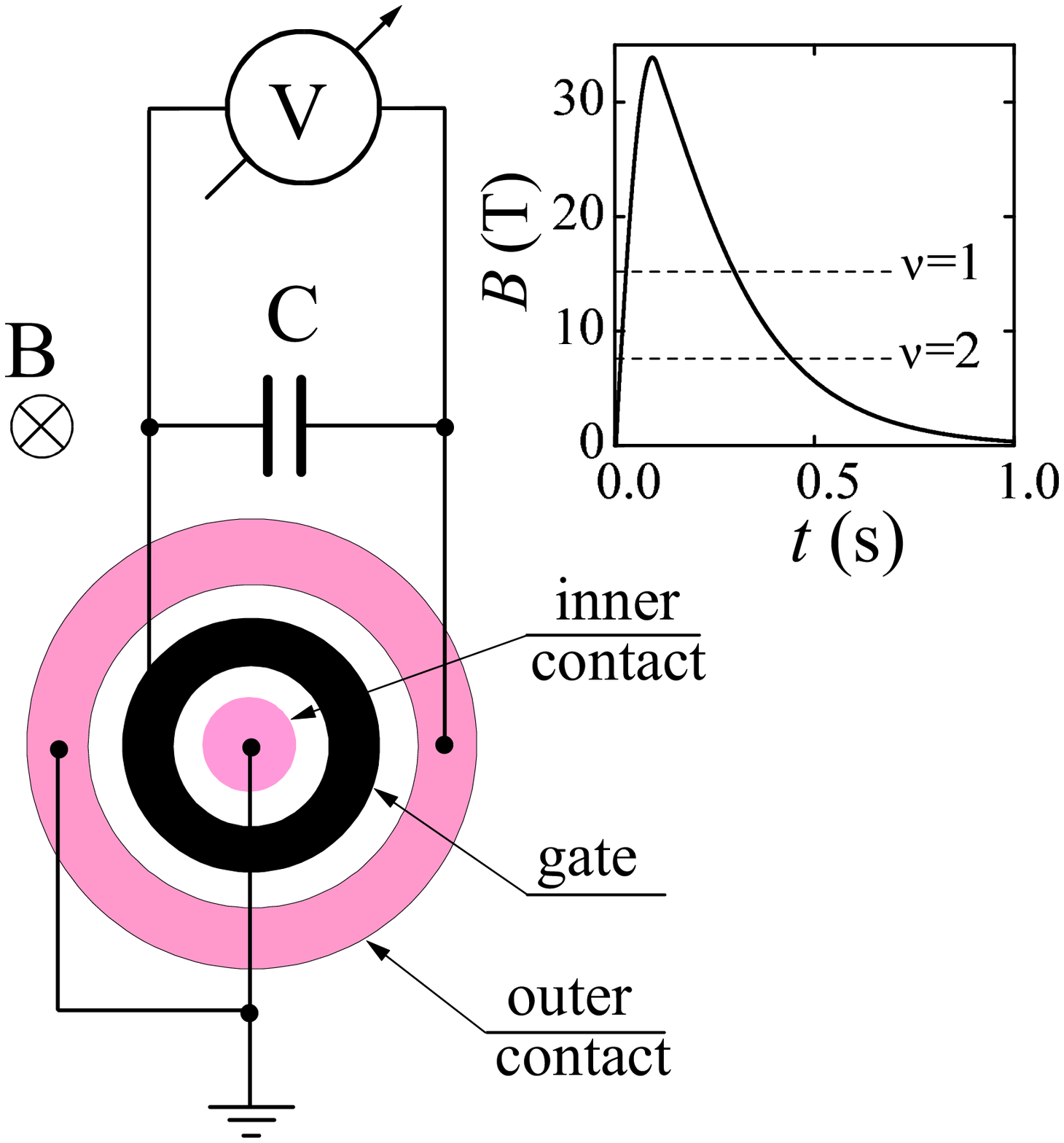,width=3.3in,bbllx=.2in,bblly=1.25in,bburx=7.25in,bbury=9.5in,angle=0}
}
\vspace{-0.9in}
\hbox{
\hspace{-0.15in}
\refstepcounter{figure}
\parbox[b]{3.4in}{\baselineskip=12pt \egtrm FIG.~\thefigure.
Schematic view of the sample and measurement circuit. The magnetic
field pulse is shown in the inset.\vspace{0.20in}
}
\label{f1}
}
}
lower) with rising and falling times of about 50~ms and 1~s,
respectively (inset to Fig.~\ref{f1}). The azimuthal electric field
induced by magnetic field sweep gives rise to an electric current
only in the radial direction if $\sigma_{xx}\rightarrow 0$. In the
experiment we study the charge, $Q$, brought out of the gated region,
which is equal to the difference between the charge exiting and
entering the gated region \cite{dol}

\begin{equation}
Q=\pi(r_{2g}^2-r_{1g}^2)\sigma_{xy}\Delta B.\label{eq2}\end{equation}
This charge induces the voltage, $V=Q/C$, across a sufficiently large
capacitance, $C$, connected in parallel to the gate, which is
measured using a preamplifier and a digitizer. The capacitance $C$
allows one to restrict the induced voltage in order to avoid the
breakdown of the dissipationless quantum Hall state \cite{rem}. The
equilibrium ($B=0$) electron density in the gated region can be
changed by using a gate bias, $V_g$. All data we show in the paper
refer to the gate voltage $V_g=0$; we have checked that for gate
voltages between 0 and $-80$~mV (the threshold voltage $V_{th}\approx
-0.3$~V), the results discussed below are not sensitive to $V_g$. In
the experiment with quasistatically changing magnetic fields (with
sweep rates in the range $(1-5)\times 10^{-3}$~T/s), we measure the
charge $Q$ using an electrometer.

Typical experimental traces of the voltage induced on the sample in a
pulsed magnetic field in the vicinity of filling factor $\nu=1$ and
$\nu=2$ are shown in Fig.~\ref{f2} for up and down sweeps. When
sweeping the magnetic field up, at small $\sigma_{xx}$ the voltage
rises linearly with $B$, in accordance with Eq.~(\ref{eq2}), until it
drops above a certain value of the magnetic field thereby signaling
the breakdown of
\vbox{
\vspace{14mm}
\hbox{
\hspace{-0.2in}
\psfig{file=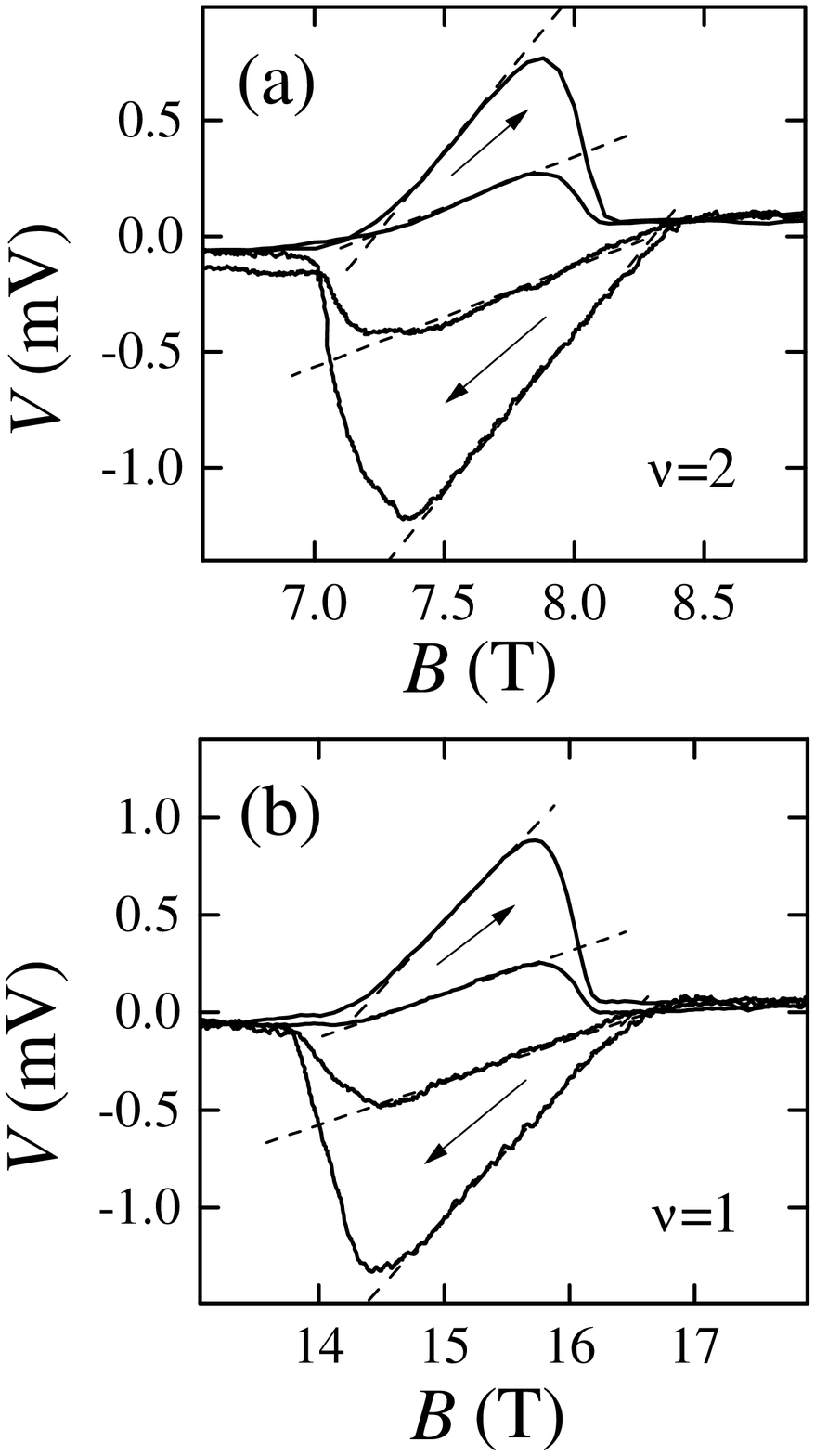,width=3.9in,bbllx=.5in,bblly=1.25in,bburx=7.25in,bbury=9.5in,angle=0}
}
\vspace{0.2in}
\hbox{
\hspace{-0.15in}
\refstepcounter{figure}
\parbox[b]{3.4in}{\baselineskip=12pt \egtrm FIG.~\thefigure.
Experimental traces of the induced voltage on one of the samples in a
pulsed magnetic field for $C=10.4$ and 32~nF at filling factor
$\nu=2$ and $\nu=1$. The expected slopes $V/\Delta B$ are shown by
dashed lines. The sweep direction is indicated by
arrows.\vspace{0.20in}
}
\label{f2}
}
}
the dissipationless quantum Hall state. On changing
the sweep direction, the voltage polarity reverses so that the up and
down traces form a hysteresis loop. The asymmetry between its top and
bottom parts is caused by larger overheating of the sample in
sweeping the field up, which leads to a more pronounced narrowing of
the quantum plateaux. A change in the background signal below and
above the hysteresis loop originates from chemical potential
oscillations \cite{dol}. Similar dependences $V(B)$ are observed also
at higher integer $\nu\le 6$ ($\nu\le 10$) for up (down) sweeps;
below we discuss the two lowest filling factors at which the observed
structures occupy the widest magnetic field intervals.

It is important that the slope in the linear interval of the
dependence $V(B)$ is in excellent agreement with the calculated one
using Eq.~(\ref{eq2}) with the quantized
\vbox{
\vspace{5mm}
\hbox{
\hspace{-0.2in}
\psfig{file=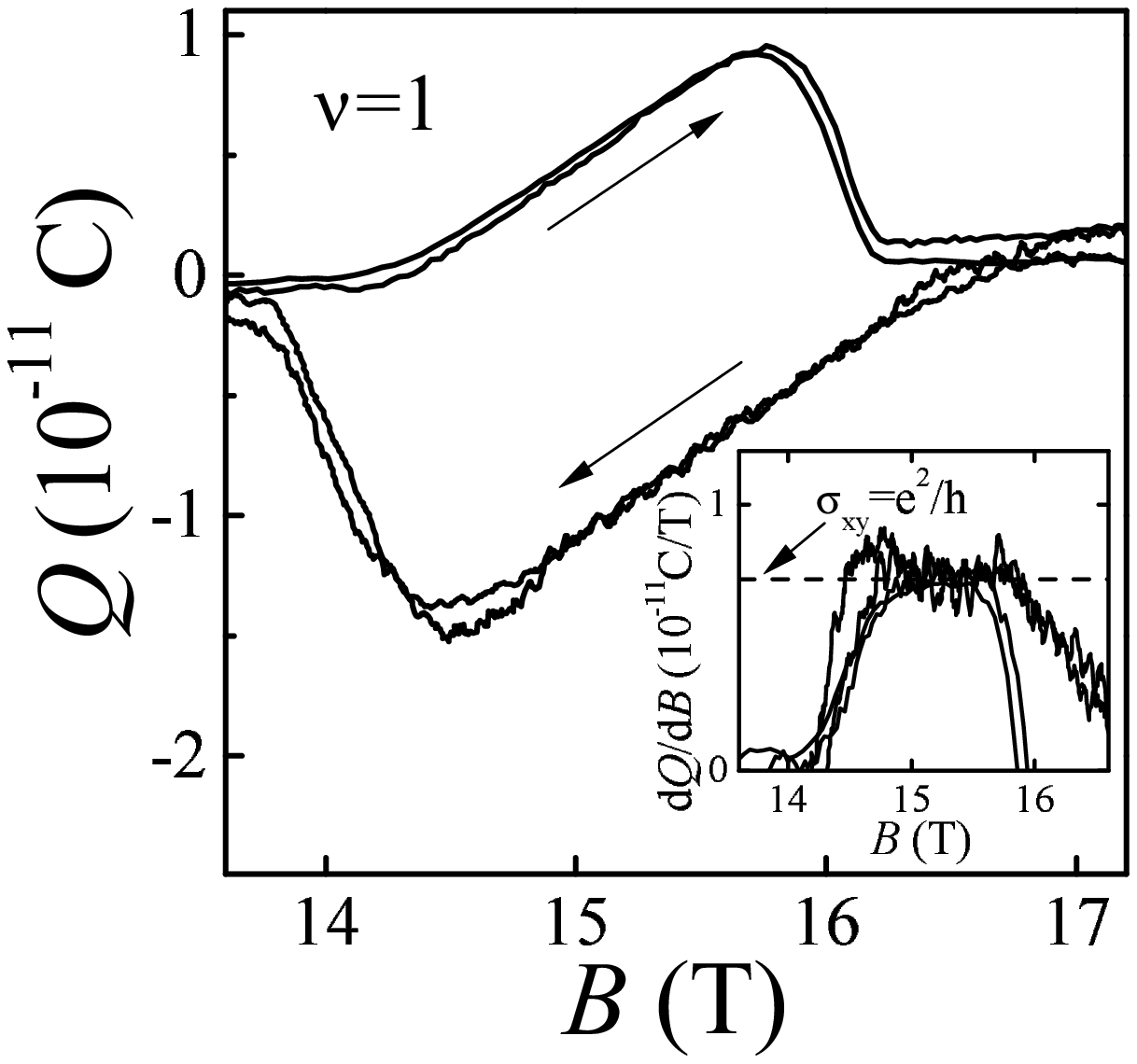,width=4.1in,bbllx=.5in,bblly=1.25in,bburx=7.25in,bbury=9.5in,angle=0}
}
\vspace{-2.2in}
\hbox{
\hspace{-0.15in}
\refstepcounter{figure}
\parbox[b]{3.4in}{\baselineskip=12pt \egtrm FIG.~\thefigure.
Charge brought out of the 2D electron system as a function of $B$ for
filling factor $\nu=1$ as obtained from the data of Fig.~\ref{f2}.
The arrows indicate the direction of magnetic field sweep. The
numerical derivative ${\rm d}Q/{\rm d}B$ is displayed in the
inset.\vspace{0.20in}
}
\label{f3}
}
}
value $\sigma_{xy}=\nu
e^2/h$, see Fig.~\ref{f2}. As the magnetic field is increased within
the linear interval of $V(B)$, electrons are brought into the 2D
system, and the electron density increases, in accordance with
Eq.~(\ref{eq2}), by $\Delta n_s=(\nu e/h)\Delta B$ (where
$\nu=1,2...$). As a result of the aligned change of magnetic field
and electron density, the filling factor remains approximately
constant: it is about 10\% larger (smaller) than the integer $\nu$
for up (down) sweep of the magnetic field (Fig.~\ref{f2}). Thus, the
observed dependences $V(B)$ yield well expressed plateaux in
$\sigma_{xy}$ as a function of filling factor.

Figure~\ref{f3} shows the corresponding dependence $Q(B)$ in a pulsed
magnetic field. As seen from the figure, within the whole hysteresis
loop, the behaviour of the charge $Q$ brought out of the 2D electron
system is independent of shunting capacitance $C$. This implies that
the observed linear $B$ dependence of the charge $Q$ is limited by a
capacitance discharge that is controlled by the dependence of
$\sigma_{xx}$ on magnetic field \cite{rem1}. The derivative ${\rm
d}Q/{\rm d}B$ yields plateaux in $\sigma_{xy}$ as a function of
magnetic field (inset to Fig.~\ref{f3}).

Typical curves of the voltage induced by the charge $Q$ in the
quasistatic measurement are displayed in Fig.~\ref{f4}. In addition
to the curves obtained by sweeping the magnetic field up and down all
way through the hysteresis loop, two more traces correspond to
reversal of the sweep direction within hysteresis loop. The expected
linear behaviour of $V$ against $B$ is shown for comparison by dashed
lines. As seen from Fig.~\ref{f4}, the intervals of the upper and
lower curves, in which $V(B)$ is linear, are narrower as compared to
the pulsed field data of Fig.~\ref{f2}.
\vbox{
\vspace{8mm}
\hbox{
\hspace{-0.2in}
\psfig{file=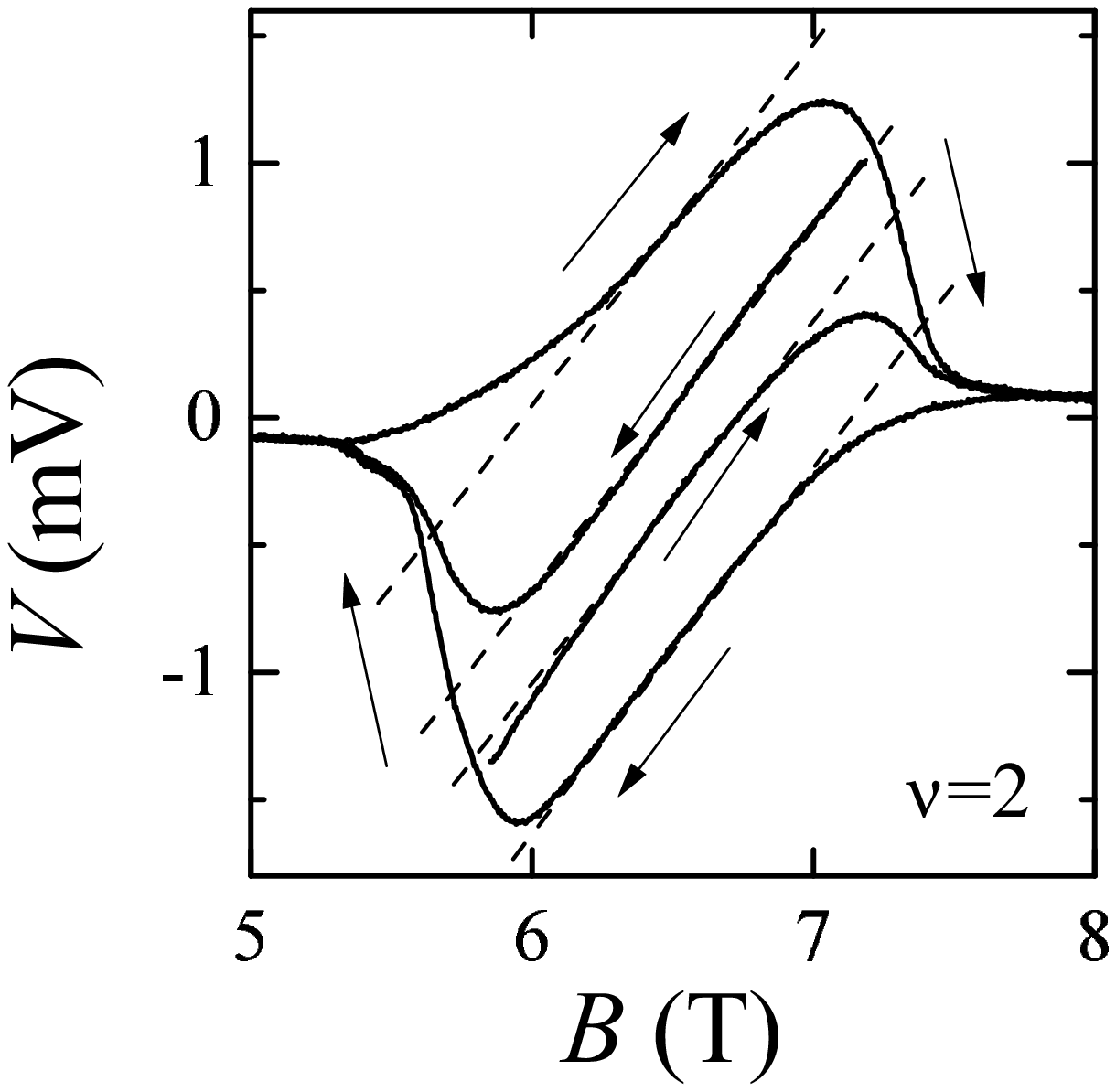,width=4.1in,bbllx=.2in,bblly=1.25in,bburx=7.25in,bbury=9.5in,angle=0}
}
\vspace{-2.1in}
\hbox{
\hspace{-0.15in}
\refstepcounter{figure}
\parbox[b]{3.4in}{\baselineskip=12pt \egtrm FIG.~\thefigure.
The induced voltage on another sample in a quasistatically changing
magnetic field as indicated by arrows; $C=30$~nF. Also shown by
dashed lines is the expected slope $V/\Delta B$.\vspace{0.20in}
}
\label{f4}
}
}
This is undoubtedly caused by
leakage currents emerging because of non-zero dissipative
conductivity $\sigma_{xx}$ whose influence is suppressed in pulsed
measurements. As a result, the accuracy of $\sigma_{xy}$ quantization
for pulsed magnetic fields turns out to be the same or even higher
than in quasistatic measurement.

One can easily see that the above mentioned adiabatic limit of the
inequality (\ref{eq1}) is not fulfilled in our experiments. This
limit corresponds to the magnetic field sweep rate $\sim 10^{-4}$~T/s
if $L=0.5$~mm, which is already an order of magnitude lower than
sweep rates in the quasistatic measurement. Moreover, the
conductivity $\sigma_{xy}$ is still found to be quantized even at
much higher sweep rates of the pulsed magnetic field, at least six
orders of magnitude beyond the estimated adiabatic limit of the
expression (\ref{eq1}). This finding unambiguously shows that the
condition of the phase coherence of the electron wave functions is
not crucial for $\sigma_{xy}$ quantization.

Apparently, our line of reasoning holds if the temperature-dependent
dephasing time, $\tau_\phi(T)$, is much larger than the phase
settling time $\tau$. The former can be evaluated from the balance
condition for thermal electron excitation to the upper quantum level
and relaxation of the excited electrons

\begin{equation}
n_0\tau_\phi^{-1}=n_0\exp(-\Delta/2k_BT)\tau_{\text{exc}}^{-1},
\label{eq3}\end{equation}
where $n_0$ is the quantum level degeneracy, $\Delta$ is the level
splitting, and $\tau_{\text{exc}}$ is the lifetime of an excited
electron. As known from optical studies (see, e.g., Ref.~\cite{qq}),
the lifetime $\tau_{\text{exc}}$ exceeds $\hbar\Delta^{-1}$ for both
spin and cyclotron splittings. Hence, we obtain $\tau_\phi >
\hbar\Delta^{-1}\exp(\Delta/2k_BT) \gg\tau$ in our experiment.

From the expression (\ref{eq1}) it follows that for our highest sweep
rates, the phase coherence of the wave functions is broken on the
length $\sim 10$~$\mu$m, which is still much larger than the magnetic
length. In other words, $\sigma_{xy}$ is found to be quantized when
the adiabatic limit is not the case for the whole sample but still
holds on macroscopic distances. Whether there exists a maximum sweep
rate of the magnetic field for $\sigma_{xy}$ to be quantized remains
to be seen.

In summary, we have measured the Hall conductivity in the arrangement
of Laughlin's gedanken experiment in pulsed magnetic fields. Well
expressed plateaux in $\sigma_{xy}$ have been observed at integer
filling factors, which is similar to the data obtained in quasistatic
measurements. Although in the pulsed magnetic field, the phase
coherence of the electron wave functions is strongly broken, the
$\sigma_{xy}$ quantization is still the case. Therefore, the
gauge-invariance-based argumentation \cite{laugh} is sufficient but
not necessary for $\sigma_{xy}$ quantization.

We gratefully acknowledge discussions with A. Gold, S.~V. Iordanskii,
V.~B. Shikin, S. Ulloa, and A. Wixforth. This work was supported by
RFBR Grants 00-02-17294 and 01-02-16424 and the Programme
"Nanostructures" from the Russian Ministry of Sciences. V.T.D.
acknowledges hospitality and support of Paul Sabatier University
during his stay in Toulouse as well as financial support of the BMBF
via a Max Planck research award.

%\begin{figure}
%\narrowtext
%\caption{\label{f1}}
%\end{figure}

%\begin{figure}
%\caption{\label{f2}}
%\end{figure}

%\begin{figure}
%\caption{\label{f3}}
%\end{figure}

%\begin{figure}
%\caption{\label{f4}}
%\end{figure}

\end{multicols}

\begin{references}
\bibitem{halp} B.~I. Halperin, Phys.\ Rev.\ B {\bf 25}, 2185 (1982).
\bibitem{buttik} M. B\"uttiker, Phys.\ Rev.\ B {\bf 38}, 9375 (1988).
\bibitem{prange} For a review, see {\it The Quantum Hall Effect},
edited by R.~E. Prange and S.~M. Girvin (Springer Verlag, Berlin,
1987).
\bibitem{laugh} R.~B. Laughlin, Phys.\ Rev.\ B\ {\bf 23}, 5632
(1981).
\bibitem{widom} A. Widom and T.~D. Clark, J.\ Phys.\ D\ {\bf 15},
L181 (1982).
\bibitem{riess} J. Riess, Europhys.\ Lett.\ {\bf 12}, 253 (1990).
\bibitem{kuchar} F. Kuchar, R. Meisels, G. Weimann, and W. Schlapp,
Phys.\ Rev.\ B\ {\bf 33}, 2965 (1986).
\bibitem{gov} S.~A. Govorkov, M.~I. Reznikov, A.~P. Senichkin, and
V.~I. Talyanskii, JETP\ Lett.\ {\bf 44}, 487 (1986).
\bibitem{yahel} E. Yahel, D. Orgad, A. Palevski, and H. Shtrikman,
Phys.\ Rev.\ Lett.\ {\bf 76}, 2149 (1996).
\bibitem{dol} V.~T. Dolgopolov, A.~A. Shashkin, N.~B. Zhitenev, S.~I.
Dorozhkin, and K. von Klitzing, Phys.\ Rev.\ B\ {\bf 46}, 12560
(1992); V.~T. Dolgopolov, A.~A. Shashkin, G.~V. Kravchenko, S.~I.
Dorozhkin, and K. von Klitzing, Phys.\ Rev.\ B\ {\bf 48}, 8480
(1993).
\bibitem{rem} Similar experiments on quasi-2D organic crystals in
Corbino geometry in pulsed magnetic fields were performed without a
shunting capacitance \cite{honold}. As a result, only a
breakdown-caused limiting behaviour was observed with no sign of
quantization.
\bibitem{honold} M.~M. Honold, N. Harrison, J. Singleton, M.-S. Nam,
S.~J. Blundell, C.~H. Mielke, M.~V. Kartsovnik, and N.~D. Kushch,
Phys.\ Rev.\ B\ {\bf 59}, R10417 (1999).
\bibitem{watts} J.~P. Watts, A. Usher, A.~J. Matthews, M. Zhu, M.
Elliott, W.~G. Herrenden-Harker, P.~R. Morris, M.~Y. Simmons, and
D.~A. Ritchie, Phys.\ Rev.\ Lett.\ {\bf 81}, 4220 (1998).
\bibitem{wiegers} S.~A.~J. Wiegers, J.~G.~S. Lok, M. Jeuken, U.
Zeitler, J.~C. Maan, and M. Henini, Phys.\ Rev.\ B\ {\bf 59}, 7323
(1999).
\bibitem{qq} I.~V. Kukushkin and V.~B. Timofeev, Adv.\ Phys.\ {\bf
45}, 147 (1996).
\bibitem{rem1} In quasistatically changing magnetic fields, a
universal breakdown curve $V(B)$ was observed \cite{dol}. The
different behaviour of the 2D electron system in the breakdown
regime, found here for pulsed magnetic fields is likely to be due to
higher electron temperatures achieved in a field pulse.
\end{references}
\end{document}